# An Ontology for the Social Determinants of Health Domain

Navya Martin Kollapally
*Department of computer science*
*New Jersey Institute of Technology*
Newark,NJ, USA
nk495@njit.edu

Yan Chen
*Department of computer information systems*
*Borough of Manhattan Community College*
New York,USA
ychen@bmcc.cuny.edu

Julia Xu
*JX consulting*
Texas, USA
julia_xu1010@yahoo.com

James Geller
*Department of computer science*
*New Jersey Institute of Technology*
Newark,NJ, USA
james.geller@njit.edu

*Abstract*— Social Determinants of Health (SDOH) are societal factors, such as where a person was born, grew up, works, lives, etc., along with socio-economic and community factors that affect an individual's health. SDOH are correlated with many clinical outcomes, hence it is desirable to record SDOH data in Electronic Health Records (EHRs). Besides storing images, text, etc., EHRs rely on coded terms available in standard ontologies and terminologies to record observations and analyses. There is a substantial amount of research on understanding the clinical impact of SDOH, ranging from screening tools to practice-based interventions. However, there is no comprehensive collection of terms for recording SDOH observations in EHRs. Our research goal is to develop an ontology that covers the terms describing SDOH. We present a prototype ontology called Social Determinant of Health Ontology (SOHO) that covers relevant concepts and IS-A relationships describing impacts and associations of social determinants. We describe the evaluation techniques that we applied to SOHO, including human experts' review and algorithmic evaluation.

*Keywords— Social determinants of health, Ontology, EHR, SDOH ontology, SNOMED CT, Protégé, BioPortal, ICD 10 CM.*

## I. INTRODUCTION

Government, healthcare organizations and community stakeholders have developed various initiatives at the county, state, and national level to identify and address the societal factors that influence health, including social determinants of health and systemic causes that lead to health inequity. The first-hand information on the impact of SDOH is identified when a patient comes for clinical or emergency visits. Hence it is necessary to record this data in an EHR. Acknowledging and understanding implicit bias can be difficult, but addressing and understanding SDOH could help advance health equity of patients under care [1]. The semantic interoperability of EHRs requires precise modelling of their clinical information. The use of an ontology can facilitates the semantic representation of clinical data and improve semantic interoperability. Data properly recorded in an EHR can be used by healthcare administrators and researchers to better understand the social needs and would support policy and payment reforms. According to Cimino, adding a "Why" in EHRs for patient care establishes relations between symptoms, physical findings, diagnostic results, differential diagnoses, and therapeutic plans [2].

Screening tools like PRAPARE (Protocol for Responding to and Assessing Patient Asset, Risk and Experience) developed by NACHC (National Association of Community Health Centers) are widely used across states and even in other countries by health care centers, primary care centers, health systems, Medicaid agencies and others [3]. PRAPARE is standardized using the LOINC [4], ICD-10-CM [5], and SNOMED CT [6] terminologies. Thus, the goal of our research is to create an ontology for SDOH with a sufficient level of granularity for practical use in a healthcare setting.

According to the World Health Organization (WHO), clinical terms coded with ICD (International Classification of Diseases) are the main basis for health recording including diseases in primary, secondary and tertiary care as well as on death certificates. ICD-11 is the latest version that has been in effect since January 2022. Using the earlier version ICD-10-CM, hospitals used Z-codes and V-codes to record SDOH terms in structured documentation in EHRs. According to the Centers for Medicare and Medicaid Services (CMS), health care professionals used Z-codes 525,987 time for Medicare fees for service beneficiaries. This corresponds to only 1.6% of the total number of beneficiaries, but it is still an increase of 6.2% over 2018 and of 17.9% over 2016 [7]. The adoption of Z-codes and V-codes was limited due to the lack of clarity on who can document the patients' social needs and the absence of an operational process for coding social demands [7]. However, the rise in term usage over the years shows an increased awareness among healthcare professionals regarding the societal impact on health.

In ICD-11, Chapter 24 is dedicated to covering the "Factors influencing health status." For example, problems associated with housing are now coded as QD71. There are different types of housing insecurity, such as living in a house with lead paint, which could cause lead poising in children. The code for this is 8B6Y for "motor neuron disorder due to lead intoxication." How could a medical professional code the situation when a child under the age of six, living in a house built before 1971, is exposed to lead poisoning, marking it as a social impact on health? There is a major gap in the codes available under this section and it is hard to rationalize why migration status, neighborhood safety, structural racism, or food insecurity are not included as factors affecting health status [8]. An insufficient level of granularity and details in coding inhibits a complete understanding of the health situation and the root causes affecting the health of patients. Our goal in this research is to provide medical practitioners with a standardized language in the form of an ontology to express social causes of health problems.

## II. BACKGROUND

EHRs have been widely used in healthcare to record demographics, vital signs, test results, immunizations, medical imaging reports, differential diagnoses, etc. EHRs rely on coded terms from underlying ontologies or terminologies to facilitate semantic interoperability [9], even though many notes appear as English text without annotations. Ontologies help with defining concepts, relationships between them, and instances that can be utilized in biomedical research. In the absence of universally agreed-upon criteria [10] which concept networks should be considered ontologies, we will refer to SOHO as an ontology, although one could argue that it is "only" a hierarchical concept graph or a taxonomy. BioPortal, maintained by the National Centre for Biomedical Ontology (NCBO), is a repository of over 900 biomedical terminologies and ontologies. An important aspect of BioPortal is ontology reuse [11]. Ontology content reuse enables a consistent representation of a domain and saves time and effort.

SDOH has a major impact on the well-being and quality of individuals' lives. The US government website Healthy People 2030 organizes SDOH into five key categories: "Economic stability," "Education access and quality," "Healthcare access and quality," "Neighborhood and built-in environment," and "Social and community context" [12]. We will discuss each of these categories below.

### A. Economic stability

Annual household income, living expenditures, socioeconomic status, housing stability, and food insecurity all have a major impact on the health and well-being of an individual. It is estimated that one in seven households with a family size of four is making less than $26,246 per annum, which is considered by the government as living in poverty [13]. In many high-priced metropolitan areas, even twice as much for a family of four would create a struggle to pay for housing, transportation, food, utilities, and clothing. Recently, the situation has been adversely impacted by Covid-19, supply chain issues, and rising inflation [14]. There are assistance programs for housing of low-income families, but they are required to pay a percentage of the rent, depending on the house they qualify for [15]. Housing instability is stressful and may impact a person's health in different ways.

Food insecurity is another area impacted heavily by the Covid-19 pandemic. Here too, there are governmental support programs such as SNAP (Supplemental Nutrition Assistance Program), which is the largest federal assisted nutrition program for low-income individuals [16]. The WIC (Special Supplemental Nutrition program for women, infants, and children) [17] assistance program was designed by the USDA (US Department of Agriculture) for women and children under 5 years struggling with poverty [17]. However, this program is not sufficient, because a family of four receives on average $477 a month, which comes out to $1.33 per person per meal. Food insecurity in the US is correlated with increased prevalence of chronic health conditions [13]. Poor workplace conditions and stress due to job insecurity may lead to drug abuse, cardiovascular diseases, anxiety disorders, etc. [18]. This has been made worse by the emergence of the "gig economy," where workers do not have any benefits and any protection. Anecdotally, some senior citizens with health insurance get coverage of drug regimens costing upwards of $50,000 a year, but do not have enough money to buy sufficient nutritious food.

In 2017 a collaboration between Food Research and the Action Center (FRAC) with the American Academy for Pediatrics attempted to identify families with food insecurity using a questionnaire that patients completed during their annual visits. A few of these families then enrolled in the SNAP program [19, 17]. Similar initiatives incorporated questionnaires and reminders in EHRs to initiate prescriptions for healthy food and on-site food banks, where all clinical options had to be properly coded in the EHRs.

### B. Educational access and quality of education

Educational inequality is driven by many factors including poverty, orphan status, substance abuse, social discrimination, low household income, etc. Education is a pathway to financial security, stable employment, and social standing. Health and longevity are adversely affected in people with lower educational achievements.

Racial and ethnic minorities mostly live in low-income neighborhoods with poorly ranked schools. De facto school segregation follows neighborhood segregation. Individuals with low levels of education reported more chronic conditions and functional disabilities [20]. Awareness of federal and state level initiatives early in the life of children helps with eliminating the different trajectories leading to educational disparities. Having terms to code "Inaccessible education due to thresholds of assistance programs" in EHRs during a pediatric visit might help the affected populations explore ways to access available government funding programs that they are unaware of.

### C. Healthcare access and quality

Inaccurate diagnoses, improper medications, unsafe clinical practices, and lack of adequate training are major factors that lead to low quality healthcare. Around seven percent of patients hospitalized experience an infection during the stay [21]. About one in ten residents in the US does not have health insurance coverage, resulting in an inability to afford medications, primary care visits, and preventive screenings [22]. Health insurance benefits are often tied to employment, thus being laid off from a job would lead to a loss of access to affordable healthcare. Many health insurance companies deny coverage of preventive services and expensive medications, which degrades the level of care that can be provided by physicians [22]. According to Laurie Zephyrin (VP of Delivery system reform at Commonwealth fund) providers armed with data can tailor their practices and help improve care. They would have a better sense of the care they are providing [23]. For providing data at multiple care sites, the EHRs must be interoperable among healthcare systems.

*D. Neighborhood and built-in environment*

The places where individuals work, live, and play have a great impact on their health and well-being [18]. There is a growing body of research showing the impact of proximity to environmental hazards and the effects that contaminations of water and air pollution have on healthy body functions [24]. The ill effects range from premature death, aggravated asthma, and non-fatal heart attacks to fetal deaths and many others [24].

A lack of access to public transportation in a neighborhood leads to delayed or missed medical appointments, delayed care, and finally delayed medication use. This leads to poorer management of illnesses and may result in severe health outcomes. Higher exposure to noise and secondhand smoke (due to blockage of a neurotransmitter) may result in greater incidences of tinnitus in adolescents and young adults [25].

*E. Social and community context*

Many people, whether they belong to a minority or not, face challenges they cannot control, including unsafe neighborhoods, discrimination, and biased policing [26]. A positive environment at work, family, and neighborhood can have a beneficial impact on the health and wellbeing of the individual. Rural residents are more vulnerable to negative societal factors such as poverty [27]. The impact of these factors includes limited access to public transportation, under-resourced schools, long commutes to obtain health care, etc.

## III. RELATED WORK

Jani et al. [28] developed a social prescribing ontology which contains 668 codes from the SNOMED CT UK version. Social prescribing is an initiative developed to address the social determinants of health. "Social prescription" is the act by which the patients are referred to community support, social events, fitness classes and social services. The coverage of this code set is intended to capture the actual interventions delivered by social prescriptions. The authors utilized the existing primary care codes to record in more detail which social prescriptions are recommended by primary care. The major categories in this ontology are limited to addiction support services, benefits signposting services, bereavement support services, dementia support services, and diabetes management support services.

Kollapally et al. [29], developed the HOME ontology for minority health equity. In that work, the authors identified gaps in standard terminologies and ontologies, such as SNOMED CT, ICD-11, NCIt, and MedDra, concerning concepts describing injuries faced by people of color within healthcare and outside of healthcare organizations. This first of its kind ontology was evaluated by standard ontology quality metrics and by a human expert. In this paper, we present an ontology based on aspects of societal factors that adversely affect around 70% of healthcare outcomes [1]. Not all patients who are negatively affected by SDOHs are minority members, thus the current work is not a superset of the HOME ontology. Environmental factors, workplace situations, educational inequality, etc. affect people of all races and ethnicities not just minority.

Patra et al. [30] extracted social determinants of health from EHRs, using NLP techniques. They utilized scholarly databases such as PubMed and ACL Anthology to identify articles related to SDOH and extracted key words from EHRs. A total of 82 publications between 2005 and 2021 were utilized for the analysis. They identified that mental health is a notable outcome associated with all the SDOH categories besides environmental factors. Emergency hospitalization was another SDOH category associated with housing, financial issues, alcohol abuse, social connection, and family abuse.

HL7 Fast Healthcare Interoperable Resources (FHIR) is an ongoing effort to create interoperable informatics tools and solutions [31]. HL7 is a set of standards used to provide guidance while using and sharing clinical data. The Gravity project initiative by SIREN (Social Intervention Research and Evaluation Service), under way since 2019, aims at developing data and exchange standards to represent patient level SDOH data across screening, diagnosis, goal setting and treatment. The Gravity work stream dashboard categorized SDOH into food insecurity, housing instability, inadequate housing, transportation insecurity, financial insecurity, material hardship, employment status, stress, etc. [32]. It provides a myriad of resources including an intervention framework, screening tools, terminology and data resources, etc., under each domain.

## IV. METHODOLOGY

In this paper, we describe the development of an ontology to cover terms for negative societal phenomena that affect clinical outcomes, the Social Determinants of Health Ontology (SOHO). Below we will describe the methods used to develop and evaluate SOHO.

*A. Concept identification*

We used the "advanced search field" on the PubMed publication website to identify relevant publications. We then reviewed and analyzed articles available in JAMIA **[33]** and other sources and extracted biomedical publications in the context of SDOH. After careful manual review of these documents, we identified major concepts used in SDOH in a clinical setting. The keywords used for extracting relevant documents were "SDOH," "poverty," "social risk," "economic instability," "food insecurity," "job insecurity", "unsafe neighborhood", "poor housing", "social and community factors affecting clinical outcomes," "poor healthcare" and "poor education." We also extracted concepts related to SDOH from the Healthy People 2030 SDOH Model [12] and County Ranking Model [34]. Having extracted the major concepts and sub-concepts, we used each concept to identify the impacts the specific class has on the health and well-being of individuals. We analyzed the sub-concepts to identify what causes would result in the concepts specified, what are the different variations, and how do they indicate an impact on clinical outcomes. We utilized manual analysis of scholarly articles published between 2011 and 2021 to identify the impacts of SDOH on the lives of individuals.

## B. Coverage in Bioportal

We used the 'class search' feature in BioPortal to determine whether the concepts we extracted are present in existing ontologies in BioPortal. The 'advanced keyword-based search' in BioPortal returned responses not limited to the exact string match, but also included synonyms and semantically similar concepts. We counted the numbers of concepts returned and manually determined how many of them were relevant. For example, "Social Determinants of Health" retrieved 43 ontologies, of which only seven were relevant in the context according to our review. We utilized the synonyms suggested for the concepts by BioPortal while designing SOHO. We used the 'find an ontology' field to search for any existing ontologies related to social determinants of health. For this purpose, we used various semantically similar concepts, such as "social determinants of health," "social survival," "societal effects," "non-clinical factors," etc. We then targeted three large, popular ontologies/terminologies, namely SNOMED CT, ICD-10 and NCIt to determine to what degree the concepts or their synonyms in SOHO are available there.

## C. Ontology integration and reuse

We integrated branches of the HOME ontology focusing on health equity into the current work. Those concepts deal with injuries, hazards and adverse incidents faced by people of color rooted in implicit bias. To enable ontology integration and reuse, we investigated BioPortal for initial established frameworks relating to SDOH. A keyword based manual search was performed in BioPortal to discover ontologies covering social determinants of health. We have not used automatic concept extraction from unstructured text in this work, due to the lack of a reference or benchmark ontology that would be needed, according to the current state-of-the-art to train the ML model for refinement or for constructing a knowledge graph [35].

## D. Ontology implementation

Protégé is the most widely used ontology editing and visualization tool. We implemented the SOHO ontology in Protégé 5.5 in OWL (Web Ontology Language) format and visualized it using OwlViz. Protégé refers to "concepts" as "classes," and allows adding annotations to classes in the user interface. The class "Thing" is predefined in Protégé and is used as the root of every ontology created with it. Protégé enables users to edit ontologies in OWL and use a reasoner to maintain consistency of ontologies. We performed consistency checking in Protégé by utilizing HermiT Version 1.4.3.456. The reasoner is based on hyper tableau calculus, which allows it to avoid nondeterministic behaviour exhibited by the tableau calculus that is utilised in FaCT++ [36] and Pellet. Figure 1 shows an excerpt of SOHO.

## E. Ontology evaluation

The HermiT reasoner in Protégé can determine whether an ontology is consistent, and it identifies subsumption relationships between classes. The reasoner determines all the inconsistent classes in the ontology. After using HermiT, we loaded the ontology with inconsistent axioms into the OntoDebug plugin [37]. OntoDebug is an interactive ontology debugging tool in Protégé. This plugin helps in identifying erroneous axioms responsible for inconsistencies. Interactive ontology debugging is implemented by iteratively stating queries in the form of wrong and correct axioms. OntoDebug will automatically recompute its diagnoses and suggest new queries. The interface provides options to create negative and positive test cases suggested by the 'Queries' tab. A few of the test cases we added are shown in Table 1. When the end user is not a domain expert, not marking the suggested axioms as positive or negative implies that its status is unknown, and this is handled in a different way in OntoDebug. Once we add the negative and positive test case they will appear under the 'Acquired test case' tab. We restart OntoDebug and evaluate the ontology again with new test cases. This is repeated until there are no more error messages.

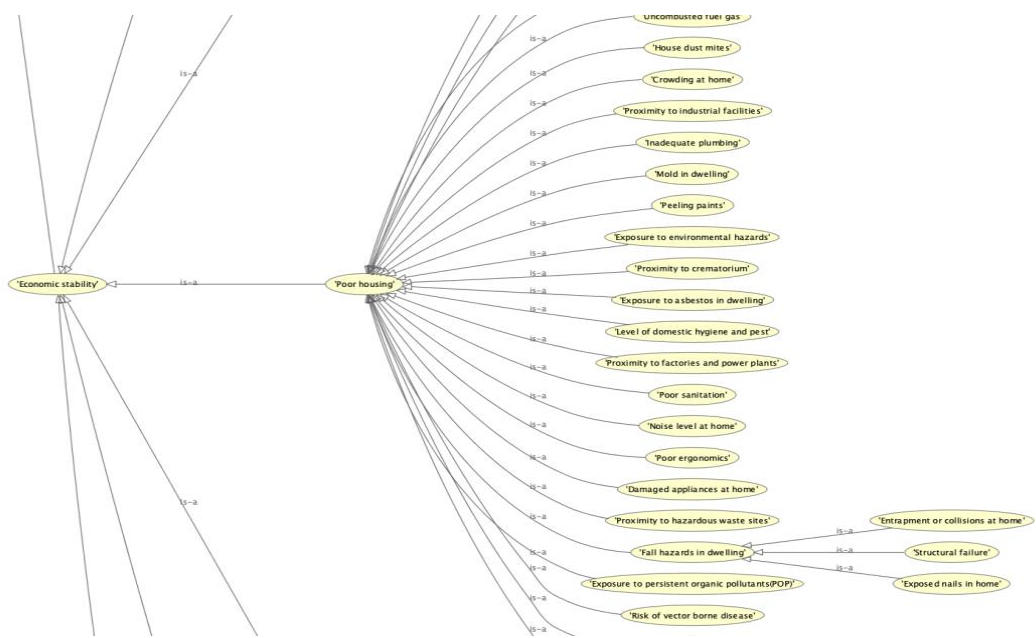

Fig. 1. Snippet from OWLViz visualisation of SOHO

TABLE 1. ACQUIRED TEST CASES FROM ONTODEBUG.

| Positive testcase | Food insecurity *SubClassOf* Economic stability |
|---|---|
| Positive testcase | Peeling paints *SubClassOf* Poor housing |
| Negative test case | Premature deaths *DisjointWith* Sicker patients |
| Negative test case | Diabetes *DisjointWith* Food_insecurity |
| Negative test case | Poor_housing *Disjointwith* Economic_stability |

*F. Human expert evaluation*

An ontology is used as a reference model, capturing knowledge about a specific domain, and providing relevant concepts and relationships. The main goal in evaluation of an ontology is to make sure that it is consistent, accurate, and has high levels of adaptability and clarity. After evaluating the SOHO ontology with the above software tools for consistency and sematic correctness, we involved two human expert evaluators with extensive experience in medical ontologies to evaluate SOHO. To understand the percentage agreement between the two evaluators we utilized Cohen's kappa coefficient. Cohen's kappa (κ) is an alternative when the overall accuracy is biased to understand the level of agreement between two evaluators [38].

For evaluation, we provided a spreadsheet with 72 randomized concept pairs. Table 2 shows a snippet from the pairs we used for evaluation. The development of SOHO followed a two-phase process. In the first phase, an initial concept hierarchy was designed, that included IS-A relationships and object properties (akin to semantic relationships or lateral relationships). This hierarchy included pairs of concepts connected by "parallel" relationships, typically an IS-A and another semantic relationship (such as Impact-Of) between the same two concepts. A preliminary evaluation review elicited feedback that (1) some of the IS-A relationships were questionable, not confirming to the criteria given in [39]; and (2) "mixing" relationships is rarely recommended, although it may be justified in some cases [40].

Therefore, in the second phase, the hierarchy was simplified to contain only IS-A links. This loss of expressivity was accepted as it brought with it a gain in precision of semantic information. Secondly, all IS-A links were reviewed by the authors with the criterion that a pair "A IS-A B" should be readable as close to a complete English sentence as possible. Whenever this was not the case, corrections were made to concept names. In many cases, desirable corrections would have resulted in very long concept names, which would lead to overwhelming ontology diagrams. Thus, as a compromise, the current concept names should be viewed as shortened noun phrases. For example, the concept pair "Noise Level at Home" IS-A "Poor Housing" (Fig 1) does not make a "good" English sentence. However, we view it as the short form of the sentence "A *Noise Level at Home Situation* is a *Poor Housing Situation*," which appears acceptable according to our English language intuition. In the Protégé implementation, the two uses of the word "situation" are implicit.

V. RESULTS

Analyzing the retrieved results from BioPortal for each concept, we found that the percentages of relevant concepts in the context of SDOH is low (see orange lines in Figure 2, based on Table 4). Even though there was a six-fold increase in SDOH-related papers in PubMed from 2011 to 2021, we could not locate an ontology covering SDOH concepts at a sufficiently fine level of granularity.

We found that 77 out of 189 (40.74%) concepts in SOHO are available in the three target ontologies. Among them SNOMED CT has the majority of terms (60 out of 189; 31.74%), followed by ICD-10-CM, which contains 31 out of 189 (16.4%) and NCIt covers 14 (7.4%) of the concepts. Note that 60+31+14=105>77, because some concepts appear in two or all three of the ontologies. Many of the major concepts in SOHO are present in terminologies such as SNOMED CT, ICD-10-CM and MeSH, as shown in Table 3. However, coverage is not consistent, and the desired level of granularity in terms of sub-concepts and semantically related concepts could not be identified using BioPortal. For example, we identified code *Z56.6* in ICD-10-CM for "physical and mental strain related to work," but codes for "stress," "drug abuse," "paranoia," "fluctuating weight," "irregular sleep cycle," etc. are not available among Z-codes in ICD-10-CM as having a semantic relationship with *Z56.6*. Using the code *Z56.6* does not provide enough granularity. Utilizing a heavy-duty tool such as *post-coordination* defeats the purpose of having readily available codes that medical practitioners can use to easily record the details of a specific case. To address this lack of granularity, we developed the SOHO ontology with the root class "Social determinants" and two children "of_health" and "of_health equity" (using a shortened form by omitting "Social Determinants_").

Each of the major concepts is related to several sub-concepts and is connected by IS-A relationships. The ontology evaluation metrics returned by Protégé are shown in Table 5. For tool-based evaluation, we first ran the HermiT reasoner on SOHO to identify problems, and then we used OntoDebug to correct inconsistent axioms. OntoDebug initially displayed the presence of nine erroneous axioms.

TABLE 2. SAMPLE OF SPREADSHEET PROVIDED FOR EXPERT EVALUATION

| Concept A | Relationship?? | Concept B | Related | Unrelated |
|---|---|---|---|---|
| Economic Stability | <--*is-a*-- ??? | Lead paints in dwelling | Yes | |
| Food insecurity | <--*is-a*-- ??? | Damaged appliances | | Yes |
| Food insecurity | <--*is-a*-- ??? | Discrimination at workplace | | Yes |
| Poor Housing | <--*is-a*-- ??? | Exposure to asbestos in dwelling | Yes | |

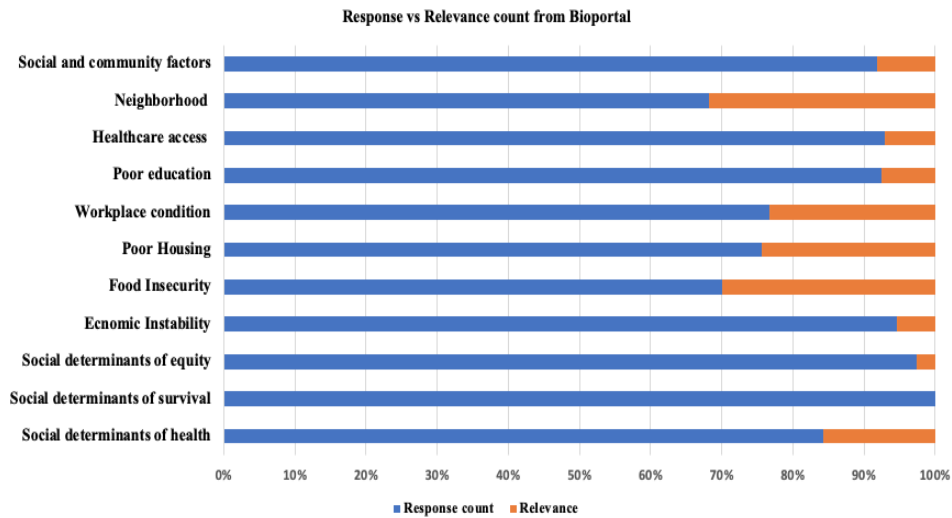

Fig .2 Response vs Relavance statistics visualisation.

TABLE 3. MAJOR CONCEPTS AND THE ONTOLOGIES CONTAINING THOSE CONCEPTS

| Concept | Ontology |
| --- | --- |
| Social determinants of health | MeSH, LOINC, IOBC, OMIT, PMA, SNOMED CT, ICD-10-CM |
| Social determinants of equity | GSS0 |
| Economic Instability | SNOMED CT, MeSH, ICD-10-CM |
| Food Insecurity | SNOMED CT, NCIT, MS, MeSH, LOINC, ICD-10-CM |
| Poor Housing | CTV3, HCDR, OCHV, IOBC, MeSH, NCIT, HL7, SNOMED CT, PMA, ICD-10-CM, ICD0 |
| Workplace condition | MeSH, OCHV, NCIT, IOBC, APAONTO, GSSO, HL7, OMRSE, ICD-10-CM, SNMI |
| Poor education | SNOMED CT, ICD-10-CM , SNMI, MeSH, PEO |
| Healthcare access | HHEAR, MEDLINEPLUS, ICD-10-CM |
| Neighborhood | LOINC, OCHV, SNOMED CT, MeSH, ICD-10-CM, HHEAR, Phenx |
| Social and community factors | MeSH, PMA, Phenx, PEO, ICD-10-CM |

After addressing the issues and updating the ontology accordingly, we reduced the number of faulty axioms in each iteration. For example, initially during the design, we defined all the concepts as disjoint from each other. Ontology repair suggested that "Poor_housing" and "Food_insecurity" had many sub-concepts in common. Hence, making them share the common concepts reduced the number of deficient axioms from 9 to 5. Similarly, by adding the repair step of making "Poor_housing" overlap with "Educational access and quality," we were able to further reduce the inconsistent axioms. Evaluating and fixing all the axioms reported by OntoDebug resolved all the inconsistencies. Thus, SOHO is a consistent ontology as per Protégé and OntoDebug.

TABLE 4. RESPONSE VS RELEVANCE COUNT OF RESULTS IN BIOPORTAL

| Concept | Response | Relevance |
| --- | --- | --- |
| Social determinants of health | 43 | 7 |
| Social determinants of survival | 37 | 0 |
| Social determinants of equity | 37 | 1 |
| Economic Instability | 35 | 3 |
| Food Insecurity | 14 | 6 |
| Poor Housing | 34 | 11 |
| Workplace condition | 33 | 11 |
| Poor education | 61 | 5 |
| Healthcare access | 39 | 3 |
| Neighborhood | 15 | 7 |
| Social and community factors | 56 | 6 |

TABLE 5. CLASS METRICS FROM PROTÉGÉ

| Metrics | Count |
| --- | --- |
| Class count | 189 |
| Axioms | 585 |
| Logical Axiom count | 207 |
| Declaration axiom count | 189 |
| Subclass of | 188 |
| Disjoint class | 19 |

Following tool-based evaluation, we switched to human expert evaluation of SOHO. The experts were provided with choices for the 66 random concept pairs, and we provided 6 correct concept pairs to the evaluators to get a flavour of SOHO. We then used a Cohen's kappa statistical calculator available online to calculate the kappa coefficient [41]. Table 6 shows the input values used to calculate Cohen's kappa. A kappa coefficient of greater than 0.4 is considered as moderate agreement and a kappa value of 1 means perfect agreement. We obtained a Cohen's kappa of 0.6363, which indicates 81.818% agreement and in turn shows that there is substantial agreement regarding SOHO between the two evaluators.

TABLE 6. VALUES FED TO COHEN'S KAPPA CALCULATOR

| Description | Count |
|---|---|
| Both evaluators agree to include | 29 |
| Both evaluators agree to exclude | 25 |
| First evaluator wants to include | 8 |
| Second evaluator wants to include | 4 |

## VI. DISCUSSION

There is growing awareness of the negative effects that nonclinical factors can have on the health and well-being of individuals. It is important to record such factors affecting patients' health in EHRs. For recording the data in EHRs, the concepts should be present in standardized medical ontologies/terminologies. Having identified gaps in available ontologies present in BioPortal, we developed an initial version of SOHO, which is available in BioPortal [42].

Associating a sick patient with the level of granularity enabled by SOHO adds to the existing knowledge base and supports the comprehensive representation of the situation of a patient in an EHR. This in turn helps healthcare providers to assist those in need according to best practices, directing the limited available resources in the best possible way to eliminate root causes of ill health. We utilized research articles from PubMed and other scholarly databases for our study and did a manual analysis of concepts, which was resource and time intensive.

### A. Limitations

One major limitation of this work is that we did not have access to actual EHR notes, as these are managed under HIPAA privacy rules. Relying more on clinical notes would have given us a better understanding of the domain in which many medical practitioners are unable to code issues in EHRs. Adding these concepts would make SOHO richer and more useful. Even though two of the authors have extensive medical background, nobody on the team is a practicing physician.

## VII. CONCLUSIONS AND FUTURE WORK

We developed a prototype version of SOHO in Protégé, with 189 classes and 585 axioms. To ensure that SOHO is consistent, adaptable, and semantically sound, SOHO was evaluated using the HermiT reasoner, and we fixed inconsistent axioms using OntoDebug. To determine whether SOHO covers domain knowledge correctly, the ontology was evaluated by two experts. The Cohen's kappa coefficient of 0.6363, indicated that both the evaluators are in substantial agreement with each other concerning the correctness of the SOHO ontology.

SOHO can be expanded and made richer with the involvement of stakeholders contributing to the knowledge from their clinical practices. Adding a Machine Learning-based analysis of clinical text would have also helped us to extract more semantically relevant concepts. SOHO could be used as a gold standard for Machine Learning-based methods to further extend it [35]. In the next phase, we will extract concepts from open publications available in healthcare platforms. We also plan to identify and extract concepts from the Observational Common Data Model (OMOP) [43], which is an observational healthcare database for studying effects of medical products.

The omission of semantic relationships such as Impact-Of in the second phase of the ontology development should not be seen as a final decision. Rather, systematic rules for which semantic relationships should be added and where they should be placed are desirable. This is left for future work. In further future work, we will extend the SDOH ontology to include concepts from the domain of Commercial Determinants of Health (CDOH) to define a comprehensive set of External Determinants of Health (EDOH).

## VIII. ACKNOWLEDGEMENTS

Research reported in this publication is supported by the National Center for Advancing Translational Sciences (NCATS), a component of the National Institute of Health (NIH) under award number UL1TR003017. The content is solely the responsibility of the authors and does not necessarily represent the official views of the National Institutes of Health.